\documentclass[conference]{IEEEtran}
\IEEEoverridecommandlockouts

\usepackage{cite}
\usepackage{url}
\usepackage{amsmath,amsfonts} 
\usepackage{float}
\usepackage{graphicx}
\usepackage{algorithm}
\usepackage{algpseudocode}
\usepackage{siunitx}
\usepackage[labelfont=bf]{caption}
\usepackage{subcaption}
\usepackage{tikz}
\usepackage{pgfplots}
\usepackage{epstopdf}
\usepackage{multirow}
\usepackage[font=scriptsize]{caption}
\usepackage{soul}
\usepackage{color, colortbl}
\usepackage{array}
\usepackage[T1]{fontenc}
\usepackage[utf8]{inputenc}
\usepackage{mathptmx}
\usepackage{listings}
\usepackage{adjustbox}
\usepackage{makecell}

\usepackage{enumitem}
\usepackage{hyperref}

\def\BibTeX{{\rm B\kern-.05em{\sc i\kern-.025em b}\kern-.08em
    T\kern-.1667em\lower.7ex\hbox{E}\kern-.125emX}}

\begin{document}

\title{Machine Learning-Based Early Detection of IoT Botnets Using Network-Edge Traffic}


\author{\IEEEauthorblockN{Ayush Kumar\IEEEauthorrefmark{1},
Mrinalini Shridhar\IEEEauthorrefmark{2},
Sahithya Swaminathan\IEEEauthorrefmark{2}, and
Teng Joon Lim\IEEEauthorrefmark{3}}
\IEEEauthorblockA{\IEEEauthorrefmark{1}Singapore University of Technology and Design, Singapore}
\IEEEauthorblockA{\IEEEauthorrefmark{2}National University of Singapore, Singapore}
\IEEEauthorblockA{\IEEEauthorrefmark{3}University of Sydney, NSW 2008, Australia}
}
			


\maketitle
\nocite{ayush-wfiot}

\begin{abstract}
In this work, we present a lightweight IoT botnet detection solution, EDIMA, which is designed to be deployed at the edge gateway installed in home networks and targets early detection of botnets prior to the launch of an attack. EDIMA includes a novel two-stage Machine Learning (ML)-based detector developed specifically for IoT bot detection at the edge gateway. The ML-based bot detector first employs ML algorithms for aggregate traffic classification and subsequently Autocorrelation Function (ACF)-based tests to detect individual bots. The EDIMA architecture also comprises a malware traffic database, a policy engine, a feature extractor and a traffic parser. Performance evaluation results show that EDIMA achieves high bot scanning and bot-CnC traffic detection accuracies with very low false positive rates. The detection performance is also shown to be robust to an increase in the number of IoT devices connected to the edge gateway where EDIMA is deployed. Further, the runtime performance analysis of a Python implementation of EDIMA deployed on a Raspberry Pi reveals low bot detection delays and low RAM consumption. EDIMA is also shown to outperform existing detection techniques for bot scanning traffic and bot-CnC server communication. 
\end{abstract}


\section{Introduction}
The Internet of Things (IoT)\cite{iotsurvey} is a network of sensing devices with low-power and limited processing capability, which exchange data with each other and/or systems (e.g., gateways, cloud servers), normally using wired (e.g., Ethernet) and wireless technologies (e.g., RFID, Zigbee, WiFi, Bluetooth, 3G/4G). IoT devices are used in a number of applications such as wearables (fitness trackers, smart watches), home automation (lighting systems, thermostats, cameras, door locks) and industrial automation (manufacturing equipment/asset management, process control, plant safety). Globally, the total number of IoT devices deployed by 2025 is estimated to reach 41.6 billion \cite{idcforecast}.

Unfortunately, hackers are increasingly targeting IoT devices using malware (malicious software) due to the following reasons\cite{iotsecsurvey1,iotsecsurvey2,iotsecsurvey3}:
\begin{itemize}
\item Many legacy IoT devices are connected to the Internet for which security updates are not installed regularly.
\item Low priority is given to security within the development cycle of IoT devices.
\item It is computationally expensive and often infeasible to implement conventional cryptography in IoT devices due to processing power and memory constraints.
\item Weak login credentials are present in many IoT devices which are either hard-coded by the manufacturer or configured by users. 
\item Sometimes, \textit{backdoors} (such as an open port) are left by IoT device manufacturers to provide remote support for the device.
\item Consumer IoT devices or the gateway through which they connect to the Internet, often do not have a firewall installed. 
\end{itemize}  

On October 21, 2016, attackers launched the biggest Distributed Denial-of-Service (DDoS) attack on record using a botnet consisting of IoT devices (IP cameras, DVRs) infected with the IoT malware \textit{Mirai}, targeting the Dyn Domain Name Service (DNS) servers \cite{miraiattack}. The Mirai source code was leaked in 2017 and since then, it has been re-used by script ``kiddies'' as well as professional hackers to build their own IoT malware. These malware are either variants of Mirai, which perform a similar brute force scan of random IP addresses for IoT devices with open TELNET ports and attempt to login using a built-in dictionary of credentials (e.g., Remaiten, Hajime), or more sophisticated ones that exploit software vulnerabilities in IoT devices to execute remote command injections (e.g., Reaper, Satori, Masuta, Linux.Darlloz, Amnesia). The various components of a typical IoT botnet and how they work with each other is explained in \cite{mirai} by taking the example of Mirai.

Bots compromised by IoT malware can be used for DDoS attacks, spamming, phishing and bitcoin mining \cite{phishspam,cryptomining1}. These attacks can cause network downtime for long periods leading to financial losses for Internet Service Providers (ISP), leakage of users' confidential data, and unauthorized exploitation of computational resources. 
Attacks originating from IoT botnets have continued unabated and have in fact, increased since the infamous 2016 Mirai DDoS attack. The author of IoT malware Satori and its improved versions (Masuta, Okiru), pleaded guilty in September 2019 to creating and managing botnets enslaving hundreds of thousands of IoT devices (including security cameras, DVR systems, home routers), launching DDoS attacks against various targets (e.g., ProxyPipe, a DDoS mitigation firm) using these botnets and running DDoS-for-hire services during the 2017-18 period \cite{satoriarrest}. Kaspersky claimed to have detected 105 million attacks on its IoT honeypots spread over the globe in the first six months of 2019, a significant increase from the 12 million attacks detected in 2018 \cite{kasperskyrep2}. On April 24, 2019, a botnet consisting of more than 400,000 IoT devices (home routers, modems, and other CPEs) launched a massive DDoS attack against a Content Distribution Network (CDN) company in the entertainment industry that lasted 13 days with peak attack flows reaching 292,000 HTTP GET/POST requests per second, making it one of the largest Layer 7 attacks handled by the DDoS mitigation firm, Imperva \cite{impervaddos}. DDoS attacks using IoT botnets continue to grow in number as well as frequency, according to \textit{The State of DDoS Weapons} report published by \textit{A10 Networks} for the fourth quarter of 2019 \cite{ddos-report-a10-netw}.

Furthermore, it is to be expected that many of the infected devices will remain so for a long time. Most IoT malware do not persist after the host device undergoes a reboot, but the device is likely to be reinfected soon by the same or another malware. Therefore, there exists a strong motivation to detect these IoT bots and manage them appropriately. It has been pointed out in \cite{trillionflaws} that it is futile to attempt to ensure that all IoT devices are secure-by-construction. Deploying traditional host-based detection and prevention mechanisms such as antivirus and firewalls for IoT devices is practically infeasible. Therefore, security mechanisms for the IoT ecosystem should be designed to be network-based rather than host-based.  

We propose a lightweight IoT botnet detection solution, EDIMA (\underline{E}arly \underline{D}etection of \underline{I}oT \underline{M}alware Scanning and CnC Communication \underline{A}ctivity), which is designed to be deployed at the edge gateway installed in home networks and targets the detection of botnets at an early stage of their evolution before they can be used for further attacks. 
EDIMA employs a two-stage detection mechanism which first uses Machine Learning (ML) algorithms for aggregate traffic classification based on bot scanning traffic patterns, and subsequently Autocorrelation Function (ACF)-based tests which leverage bot-CnC messaging characteristics at the per-device traffic level to detect individual bots. We only target IoT botnets with centralized Command-and-Control (CnC) architecture in this work. Historically, centralized botnets have been more widely deployed in real-world networks (PC-based as well as IoT), compared to Peer-to-Peer (P2P) botnets \cite{botnet-arch} as they are easier to set up and manage, though P2P botnets that avoid the single point-of-failure in centralized botnets are growing in number. 

EDIMA also includes a database that stores malware traffic packet captures, a policy engine which determines the further course of action if gateway traffic has been classified as malicious and a traffic parser/sub-sampler which sorts the aggregate gateway traffic and optionally sub-samples the traffic at a specified rate. EDIMA can be deployed both on physical edge gateways or as Network Function Virtualization (NFV) functions at the customer premises in an SDN-NFV based network architecture, where SDN stands for Software-Defined Networking. EDIMA can also be deployed alongside an intrusion detection system at the gateway. 

\textit{Justification For ML-based Detection}: While TELNET port scanning can be countered by deploying a firewall or intrusion detection system (at the edge gateway) which blocks incoming/outgoing TELNET traffic, malware that exploit software vulnerabilities targeting application protocols such as HTTP, SOAP, PHP, etc., are difficult to block using a rule-based approach because these protocols are used by legitimate traffic as well. Furthermore, deep-packet inspection IDS is not a practically viable solution for consumer IoT devices (connected to ISP networks through gateways) which are the focus of our work. It is computationally too expensive to be deployed at such gateways and requires specialized hardware. Hence, we propose to use an ML-based approach instead for detecting those sophisticated IoT malware. Utilizing carefully selected malware traffic features independent of application protocols, ML algorithms can be trained to distinguish between benign and malicious traffic (containing malware-generated packets).


EDIMA targets IoT botnets at an early stage of their evolution, when they are scanning for and infecting vulnerable devices, as the scanning and propagation phase of the botnet life-cycle stretches over several months \cite{miraiusenix}. This means that IoT bots can be detected and isolated long before they can be used for attacks such as DDoS. If a botnet has reached substantial size and started launching attacks already, they can be detected using existing methods in literature and industry to defend against such attacks.

The major contributions of this paper are listed below:
\begin{enumerate}
\item We characterize the scanning and bot-CnC server communication traffic patterns for IoT malware. 
\item We propose a lightweight IoT botnet detection solution, EDIMA, employing a two-stage detection mechanism which first uses ML techniques for aggregate traffic classification at the edge gateway and subsequently, ACF-based tests to detect individual bots. We also evaluate EDIMA's detection performance comprehensively. 
\item We implement EDIMA using Python, deploy it on a Raspberry Pi and evaluate its runtime performance. 
\end{enumerate} 

A preliminary version of this work was presented in \cite{ayush-wfiot}. The following points illustrate the ways in which we have expanded upon our previous work:
\begin{itemize}
\item We have refined the architecture for our proposed IoT botnet detection solution, EDIMA, and explained the functions of its modules in greater detail.
\item New features have been added for detection of scanning activity in aggregate gateway traffic.
\item We have introduced a novel, second stage for individual bot detection after the first stage of aggregate gateway traffic classification.
\item The performance evaluation testbed has been expanded to include more IoT devices.
\item Instead of emulating the behavior of IoT malware to generate malicious traffic training data as done in our previous work, we use real, live malware samples.
\item Unlike our previous work, we have followed all the required steps for data processing before applying ML classifiers.
\item We have done a performance comparison with few existing works in literature.
\item We have also evaluated the scalability performance and runtime performance of a software implementation of EDIMA.   
\end{itemize}  

The rest of this paper is organized as follows. In Section \ref{literature}, we review a few prominent works on detecting botnets exploiting CnC communication features, intrusion and anomaly detection systems for IoT, detecting attacks originating from IoT botnets, and IoT device traffic fingerprinting. Subsequently, in Section \ref{edimaarch}, our proposed IoT botnet detection solution, EDIMA is presented while in Section \ref{edimacompdesc}, the operation of EDIMA's modules is described in more detail. Finally, in Section \ref{perfeval}, we present the aggregate traffic classification and bot detection performance results for EDIMA, evaluate the runtime performance of its Python implementation and compare EDIMA's performance with previous works.

\section{Related Work}
\label{literature}
Botnet detection using scanning traffic features has not been adequately addressed in the literature. In particular, there are no existing works on IoT botnet detection using scanning traffic features. However, there have been several prominent works on detecting PC-based botnets using their CnC communication features. In \cite{ircbotnet}, ML-based techniques for CnC traffic detection in Internet Relay Chat (IRC) botnets have been presented which first distinguish between IRC and non-IRC traffic and then between bot and real IRC traffic. This approach of classifying traffic into IRC and non-IRC to detect IRC-based botnets has also been used in other works\cite{rishibot,srutibot,hotbots2}. A \textit{bot infection dialog model} has been built in Bothunter\cite{bothunter} based on which the authors have constructed three bot-specific sensors and performed correlation between inbound intrusion/scan alarms and the infection dialog model to generate a consolidated report. In Botsniffer\cite{botsniffer}, the authors have captured spatio-temporal similarities between bots in a botnet (in terms of bot-CnC coordinated activities) from network traffic and utilized them for botnet detection in a local area network. BotMiner\cite{botminer} has proposed a botnet detection system in which similar CnC communication traffic and similar malicious activity traffic are clustered, and subsequently cross cluster correlation is performed to detect bots in a monitored network. A number of other works have used clustering of flows with similar characteristics to detect botnets\cite{strayer1,strayer2,dimvabot}. 

A few research works have also dealt with intrusion detection systems for IoT. In \cite{heimdall}, the authors have presented a whitelist-based intrusion detection system for IoT devices, \textit{Heimdall}, which uses dynamic profile learning and runs on IoT gateway routers. An intrusion detection model for IoT backbone networks has been proposed in \cite{tldrtc} which employs two-layer dimension reduction and later, two-tier classification techniques for User-to-Root (U2R) and Remote-to-Local attack (R2L) detection. Anthi et al. \cite{smart-home-id-iot} have proposed a three layer intrusion detection system targeted at smart home IoT devices, which detects network security attacks through  normal behaviour profiling of IoT devices connected to the network, malicious packets' identification during the attack and subsequent attack classification. 

Of late, the research community has been interested in IoT botnet and attack detection with a number of papers addressing these problems. There have been a few works on anomaly detection systems for IoT that build normal communication profiles for IoT devices, i.e., the profiles are not expected to change much over a long period of time (DEFT\cite{deft}). Any deviation from those profiles is flagged as anomalous traffic (D\"{I}oT \cite{diot}, NETRA \cite{netra}). Hamza et al. \cite{mudprofile} have proposed a tool to automatically generate Manufacturer Usage Description (MUD) profiles for a number of consumer IoT devices. Using those MUD profiles, the authors have derived flow rules to be configured on SDN switches, identified their limitations in detecting network attacks on IoT devices and addressed those limitations by proposing an anomaly detection-based approach used in conjunction with an off-the-shelf IDS \cite{mud-SDN-ID}.

In \cite{nbaiot}, an anomaly detection technique built using deep-autoencoders has been used to detect attacks launched from IoT botnets. Statistical features have been extracted from behavioural snapshots of normal IoT device traffic captures which are used to train deep learning-based autoencoders (for each IoT device). Subsequently, the reconstruction error for traffic observations is compared with a threshold for normal-anomalous classification. Following a slightly different approach, Kitsune \cite{kitsune} proposes an online unsupervised anomaly detection system based on an ensemble of autoencoders which is lightweight in terms of memory footprint and meant to be deployed on network gateways and routers to detect attacks on the local network.

Our work addresses few important gaps in the IoT botnet detection literature. First, botnet detection techniques using CnC communication features \cite{ircbotnet, bothunter,botsniffer,botminer} are designed for PC-only networks whereas our botnet detection solution is designed specifically for IoT networks. IoT networks pose a number of \textit{unique research challenges} as compared to PC-only networks:
\begin{itemize}
\item In a typical IoT network, multiple IoT/non-IoT devices are connected to the Internet through a gateway which may be a wired/wireless access point (also acting as a modem) or a cellular base-station whereas PC-only networks consist of mostly PCs connected to the Internet directly through modems.
\item PC traffic is more user driven and diversified since a number of different user-level applications, e.g., browsing, email, word processor, run on a PC simultaneously and generate traffic based on user behaviour. IoT devices, on the other hand, are always-connected, run one or two applications only, and send data to servers regularly without user intervention. As a result, their traffic has discernible patterns \cite{deft} different from PC traffic. 
\end{itemize}
Moreover, the PC-only botnet detection schemes proposed in \cite{ircbotnet, bothunter,botsniffer,botminer} are designed for enterprise networks whereas our work targets home networks. Also, the PC-only botnet detection solutions are based on detection of bot-CnC traffic only whereas our work is based on detection of scanning activity first and subsequently bot-CnC communication for a more accurate bot detection scheme. 

Second, our focus is on detecting botnets irrespective of the type of CnC communication protocol used (e.g., IRC, SMTP), unlike \cite{ircbotnet}. Third, in this work, we aim to detect individual bots (instead of detecting the network of bots in a botnet directly) which can then be used towards detecting the botnet. This can be done, for example, by correlating the positions of detected bots within a network, or by using techniques based on clustering algorithms and cross-cluster correlation proposed in existing works on PC-only botnet detection \cite{botsniffer,botminer}.

Fourth, our aim is to detect IoT bots long before they are used to launch attacks, during the scanning/infection phase, unlike \cite{nbaiot,diot}. Finally, instead of using normal IoT devices' traffic fingerprinting \cite{deft} and subsequent anomaly detection \cite{diot, netra}, we follow the approach of misuse-detection where we detect IoT bots by utilizing certain distinguishing features of bot scanning traffic and bot-CnC communication traffic in succession. The former approach suffers from limitations such as an inability to narrow down the cause of anomaly, possibility of misclassification of a bot as a legitimate device type, limited testing against evasive malware which can easily trick the anomaly detection system by manipulating device traffic (because fingerprinting is performed at the device-level), etc. 

Kitsune\cite{kitsune} suffers from a similar limitation which is common to most anomaly detection systems, i.e., an inability to point out the source of anomaly. Further, it has been tested with one only IoT malware, Mirai, which is relatively simple to detect and it is difficult to say whether Kitsune will perform well against other IoT malware in the real-world. Again, Kitsune is more focused on attack detection rather than early-stage detection of botnets which is the focus of our work. Though Kitsune advocates online algorithms as it is focused on detecting attacks, for the use-case of early-stage detection of IoT botnets, offline algorithms are deemed to be sufficient as botnets take time to evolve before they are used for attacks. As expected, EDIMA is not free from limitations too, which is explained in Section \ref{limitations}. Rather, a combined approach consisting of EDIMA and IoT device fingerprinting plus anomaly detection is advocated for a more robust IoT botnet detection performance.


\section{EDIMA Architecture}
\label{edimaarch}
\begin{figure}[h]
\centering
\includegraphics[scale=0.3]{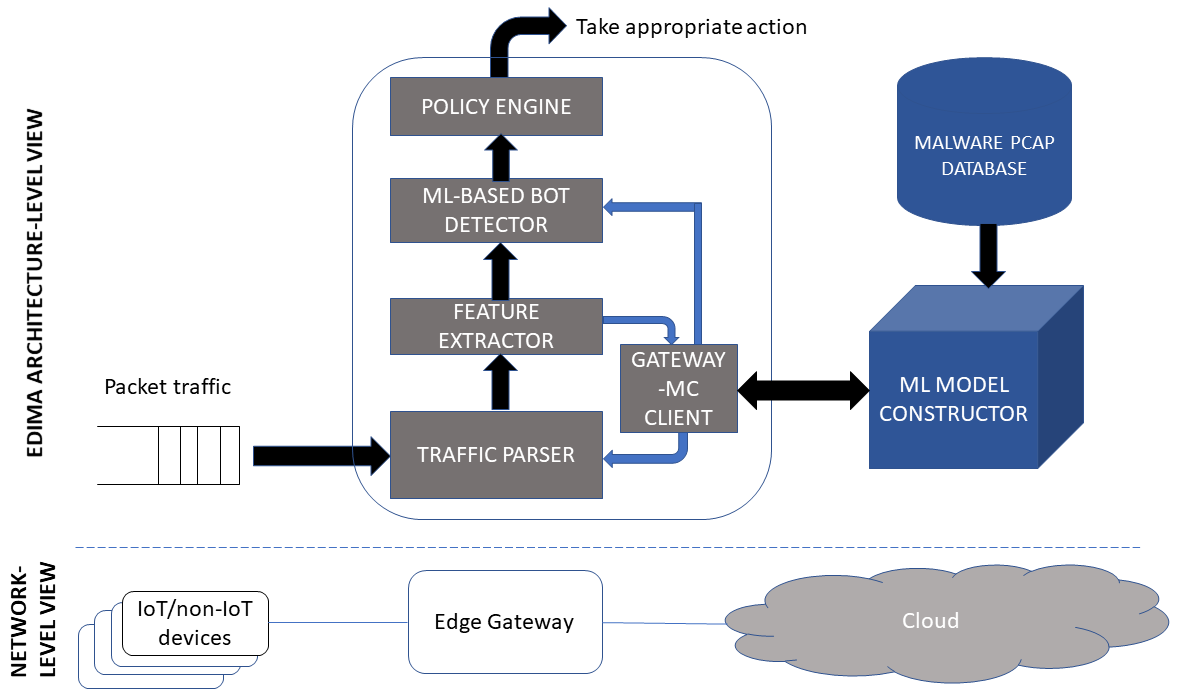}
\caption{EDIMA Architecture}
\label{EDIMA-arch}
\end{figure}

EDIMA is designed to have a modular architecture, as shown in Fig. \ref{EDIMA-arch}, with the following components:
\begin{enumerate}
\item \textbf{Feature Extractor}:
This module extracts features from the aggregate traffic at the gateway. These features are then forwarded to the ML-based Bot Detector (MBD), which is described below, for classification during the execution phase. The Feature Extractor (FE) also sends features extracted from the aggregate traffic to the ML Model Constructor (MC), which is also described below, during the training phase. The feature extraction process is explained in Section \ref{classifierdetails}.

\item \textbf{ML-based Bot Detector}:
This is a 2-stage module with the first stage being a \textit{coarse-grained} one that classifies the aggregate traffic samples using the features obtained from FE and the ML model trained and forwarded by the MC. Depending on the result of the classification, the second \textit{fine-grained} stage attempts to identify the infected IoT device(s) from the set of devices connected to the gateway. More details about the classification operation are given in Section \ref{classifierdetails} while the bot detection process is described in Section \ref{botidentifier}.

\item \textbf{Traffic Parser/Sub-sampler}:
The traffic parser (TP) sorts the combined gateway traffic into traffic sessions. During the bootstrap (training) phase of EDIMA, it also helps replay malware traffic samples along with normal traffic to generate malicious traffic samples. We also propose an option for traffic sub-sampling as introduced in \cite{ayush-ficc}. Under this option, the packet traffic from IoT devices is sampled across the devices and the sub-sampled traffic is then forwarded to the FE. This operation would help reduce the storage overhead but there is a trade-off in terms of detection delay. 

\item \textbf{Malware PCAP Database}:
This frequently updated database stores malware traffic \textit{pcap} files captured from private and professional honeypots targeted at IoT malware. We envisage a community of authorized security researchers who will collect the above \textit{pcap} files and upload them to the database through an online interface. The files will be anonymized before uploading to ensure user privacy. The \textit{pcap} files are retrieved by the MC when required.  

\item \textbf{ML Model Constructor}:
The ML model used for classifying edge gateway traffic is trained by this module. We assume a publish-subscribe model where multiple gateways subscribe to an MC. A separate ML model is trained for each gateway for optimal performance. Whenever a gateway comes online, it registers with the MC. Malicious traffic samples from the \textit{Malware PCAP Database} (mDB) are sent to the gateway to generate malicious aggregate traffic. The feature vectors extracted from benign (normal traffic with no malicious scanning packets) and malicious aggregate traffic are subsequently sent by a gateway's FE to the MC. The extracted features are used to train a supervised ML classifier which is then published to the gateway's MBD. The number of data samples required for training depends on a number of factors such as the model hyper-parameters, the desired classification accuracy/precision/training time, etc. The MC interacts with a gateway through a \textit{Gateway-MC client}. 

The gateway's ML model may have to be re-trained, e.g., when the ML model classifies with a low probability, or when the training dataset is updated, and the new model is published to the gateway. By stating that the existing ML model may classify with low probability, we mean that the classification probability falls below a pre-defined threshold (e.g., $80\%$). 

Further, the training data used to train the ML model in MC needs to be updated regularly. The training data representing benign traffic requires periodic updates for various reasons: a new IoT device may have joined the gateway, firmware update to a connected device, etc. If bot(s) have not been detected in the last traffic session, the FE extracts features from the subsequent benign aggregate traffic sessions and forwards them to the MC through the gateway-MC client. Similarly, the training data representing malicious traffic needs periodic updates due to addition of newly discovered IoT malware to mDB. The MC retrieves the new malware pcap files from mDB and publishes them to the TP which replays those malicious pcap files at the gateway. The malicious traffic so obtained is forwarded to the FE which extracts the malicious traffic features and sends them to the MC through the gateway-MC client. The period for benign and malicious training data set updates should be set separately as the frequency of changes in benign traffic is expected to be lower than in malicious traffic (due to new IoT malware being created and discovered almost every day).



\item \textbf{Policy Engine}:
The policy engine (PE) consists of a list of policies defined by the network administrator, which determine the course of actions to be taken once an IoT device connected to the edge gateway has been detected as a bot. For example, the entire traffic originating from bots can be blocked by the network administrator. The underlying IoT devices can be brought back online once they have been cleared of the malware. The details of this module are discussed in Section \ref{policyengine}.

\end{enumerate}
EDIMA's architecture is designed to be scalable in the sense that, as the number of gateways in a network increases, more MCs can be deployed with each MC catering to a group of gateways. As the MC is implemented in the cloud, the amount of resources allocated to it can be increased easily, depending on the requirement. The cloud service using which MC and mDB are implemented should be secure by industry standards. This can be ensured by employing cloud services offered by well-known commercial cloud service providers (e.g., Amazon, Microsoft, Google).

\begin{figure*}
\centering
\includegraphics[scale=0.25]{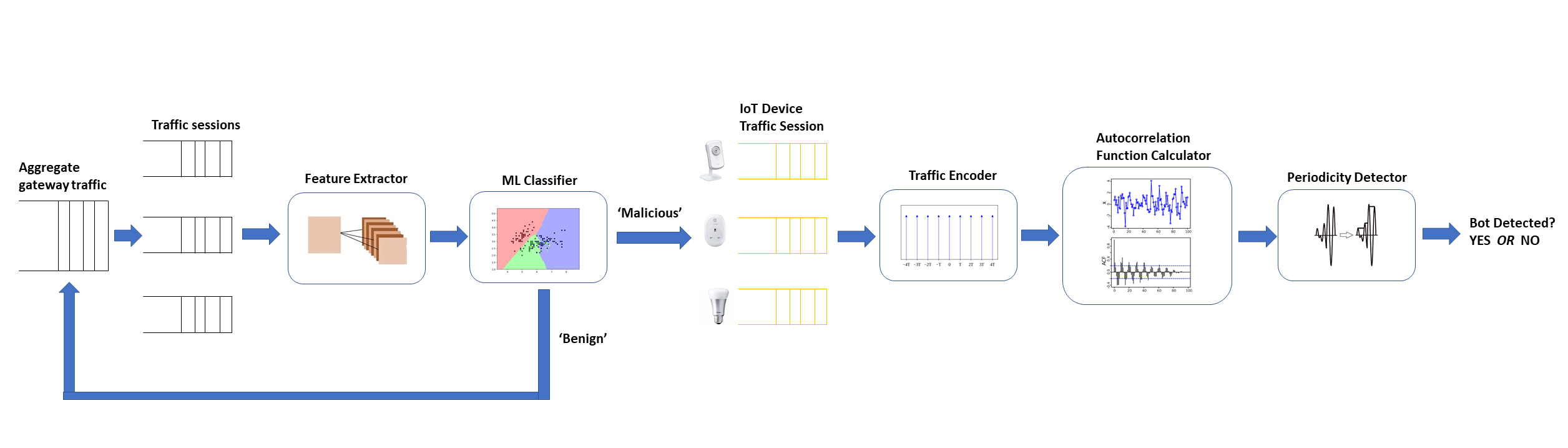}
\caption{Combined workflow of Traffic Aggregator, Feature Extractor and ML-based Bot Detector}
\label{EDIMA-bot-detect}
\end{figure*} 

\section{Description of EDIMA's Components}
\label{edimacompdesc}
\subsection{Detection of Scanning Activity in Aggregate Gateway Traffic}
\label{classifierdetails}
The first coarse-grained stage of the ML-based Bot Detector performs classification on aggregate gateway traffic rather than per-device traffic. According to Basil et al. \cite{botnetdetwalker}, one needs to keep track of less fine-grained traffic details while monitoring aggregate traffic behaviour, compared to inspecting individual packets. This might result in a lower detection accuracy and false positives, but offers several advantages such as lower computational load and faster analysis, thus contributing to higher scalability. Moreover, using less fine-grained details in the characterization of malicious traffic results in a detection scheme which is more resilient to evasion. The attacker would have to compromise the efficiency of the malicious activity to evade detection.

The aggregate gateway traffic is assumed to belong to either of the following classes: \textit{benign} or \textit{malicious}. Here, benign traffic refers to gateway traffic which does not consist of bot scanning packets whereas malicious traffic refers to gateway traffic that does. The gateway traffic is captured in the form of traffic sessions which are defined statically as the set of ingress/egress packets at a network interface over a fixed time interval. Further, the classification algorithm is applied on these traffic sessions rather than individual packets. This is because per-packet classification is computationally much more expensive and does not yield any significant benefits over per-session classification.

In a traffic session, we extract features from TCP packet headers only and not the payloads. Though payloads may be quite helpful for classification, it is possible that are encrypted (e.g., HTTPS traffic). By not depending on payload information, we ensure that our aggregate traffic classification method works on unencrypted as well as encrypted traffic. Below are the steps for gateway-level traffic classification:
\begin{enumerate}
\item Filter each gateway traffic session to include only TCP packets.
\item Perform feature extraction for each traffic session.
\item Apply the trained ML classifier on the extracted feature vectors and classify the corresponding sessions.  
\end{enumerate} 


We have carefully identified the following eight features for ML classification which are relevant for detecting bot scanning traffic (instead of using generic network traffic features):
\begin{enumerate}
\item Number of unique TCP SYN destination IP addresses
\item Number of packets per unique destination IP address
	\begin{enumerate}
	\item maximum
	\item minimum
	\item mean
	\end{enumerate}
\item Number of TCP half-open connections
\item TCP packet length
	\begin{enumerate}
	\item maximum
	\item minimum
	\item mean
	\end{enumerate}
\end{enumerate}
The intuition behind selecting the first feature is that most IoT bots generate random IP addresses and send malicious packets to them. Hence, the number of unique destination IP addresses in the presence of malware-induced scanning traffic will be significantly higher than benign traffic. The second set of features was selected because malware typically send very few scanning packets to the same IP address, in order to cover as many devices as possible. The third feature seeks to exploit the fact that while scanning, a bot opens multiple TCP connections to random IP addresses, the majority of which do not respond and so the connections remain half-open. Finally, compared to packet length in benign TCP connections, malware scanning packet lengths are generally shorter, making the fourth set of features useful for classification.

\subsection{Detection of Individual Bots using Bot-CnC Communication Patterns}
\label{botidentifier}
Once the aggregate traffic at an edge gateway has been classified as \textit{malicious}, the second fine-grained stage of the ML-based bot detector attempts to detect the underlying bots by checking the ingress/egress traffic from each IoT device for the presence of bot-CnC communication patterns. The set of IoT devices is sorted according to their IP addresses. For the sake of efficiency, we propose that the set be divided into two halves, and each half should be checked for bots in parallel. Here, we are not making any assumptions about the likely locations of the infected devices in the set. The complete bot detection algorithm is shown in Algorithm \ref{A1}. The degree of parallelization in our algorithm can be increased depending on the delay requirements.

\begin{algorithm}[h]
	\centering	
	\caption{\scriptsize{Detect\_IoT\_bots (SET\_DEVS)}}
	\label{A1}
	\begin{algorithmic}[1]
		\State \textbf{INPUT}: SET\_DEVS (Set of all IoT devices behind gateway)
		\State Initialize \textit{Thread1}, \textit{Thread2}, \textit{InfectedDevList}
		\State Thread1:
		\For {$devindR=1$ to $\frac{n(SET\_DEVS)}{2}$}
			\If {\textsf{Detect\_bot\_cnc\_comm} (devindR) $=$ PERIOD\_DETECTED}
				\State \textsf{Append} (InfectedDevList, devindR)
			\EndIf
		\EndFor
		\State Thread2:
		\For {$devindL=n(SET\_DEVS)$ to $\frac{n(SET\_DEVS)}{2}+1$}
			\If {\textsf{Detect\_bot\_cnc\_comm} (devindL) $=$ PERIOD\_DETECTED}
				\State \textsf{Append} (InfectedDevList, devindL)
			\EndIf
		\EndFor
		\State return (InfectedDevList)
	\end{algorithmic}
\end{algorithm}

\textbf{Periodicity Detection in Bot-CnC Communication} In most existing IoT botnets, including the Mirai-variants \cite{mirai}, there is a periodic exchange of TCP messages ([PSH, ACK], [ACK]) or UDP messages between the bot and the CnC server. This assumption is supported by our empirical analysis of a number of IoT malware samples belonging to different malware categories (Appendix \ref{iot-malw-categ}) as detailed in Section \ref{datacollect}, all of which exhibited the above bot-CnC communication behaviour. We would like to point out here that it is practically infeasible to analyse all possible IoT malware and it is a general practice followed in botnet detection literature to test with malware samples belonging to a few distinct malware families.

It seems fairly straightforward that to detect the presence of bot-CnC communication, we filter the traffic from a potential bot for UDP packets or TCP packets (with PSH and ACK flags \textit{ON}) and check for periodicity in the transmission times of filtered packets. However, IoT devices also send legitimate application data to a cloud server periodically. Filtering the device traffic for TCP/UDP packets alone may lead to false positives, i.e., device-cloud server traffic may be mistaken for bot-CnC server messaging traffic. 

Therefore, we further exclude IoT application data packets from our analysis using appropriate packet capture filters. It has been observed from the packet captures of uninfected commercial IoT devices and IoT malware running on embedded devices (Section \ref{datacollect}) that as opposed to the legitimate application data packets from IoT devices, bot-CnC communication TCP/UDP packets have TCP payload consisting of a few bytes only. Hence, we build a packet capture filter to exclude packets consisting of TCP payload greater than a few bytes (e.g., 5 or 10) from our analysis to obtain TCP/UDP packets related to bot-CnC communication. Even then, we may be left with TCP/UDP packets sent by desktops or phones to a server (e.g., user requesting for a web page through HTTP GET), though they are not periodic and can be considered as noise packets. 

Subsequently, we sample and encode the filtered packets to produce a uniformly sampled discrete-time signal using the approach suggested in \cite{signal-proc-netw-traffic} for further signal processing. Let the time of arrival of filtered packets be denoted as $t_j$, where $j = 0,\dots,N-1$ and $N$ is the number of packets. Then, the encoding function is represented as:
\begin{equation}
e(i) =
	\begin{cases}
 		1, & \text{if}\ iT < t_j < (i+1)T \\
		0, & \text{otherwise}
	\end{cases}
\end{equation} 
where $i = 1,2,\dots,K$ and $T$ is the sampling time interval.

To detect periodicity in time series data, the usual approach followed in literature \cite{priestley} is to calculate the power spectral density (PSD) of the time series (which can be estimated by a periodogram). Then, a statistical hypothesis test (e.g., Schuster's test, Walker's test) is used to compare a statistic computed from the periodogram's frequency components against a threshold to decide between the null hypothesis $H_0$ (underlying time series is not periodic) and the alternate hypothesis $H_1$ (underlying time series is periodic). The threshold is calculated from the asymptotic probability distribution (when $K \to \infty$) of the test statistic. 

However, in a real-world implementation, we cannot wait to collect so many samples (for the test statistic to converge to its asymptotic distribution) as that would lead to a large delay in bot detection. This is because the typical time interval between consecutive bot-CnC messages is of the order of few minutes. To collect a statistically significant  number of samples, for example $500$, assuming the bot-CnC messaging interval as $1$ minute, we would have to capture an IoT device's traffic for at least $500$ minutes $\approx 8$ hours. Even the Fisher's test, which gives an exact probability distribution expression for calculating the comparison threshold, exhibits poor statistical power \cite{fisher-test-short} when applied to short length time series (less than 40-50 time points).

In this work, we use the autocorrelation function (ACF). ACF can be used to reveal hidden periodicity in a signal even if it is corrupted by noise. For short length time series signal such as desired in our use case (bot detection), calculating ACF directly is not very computationally expensive. The unbiased ACF of discrete time signal $e(i)$ (with mean $\overline{e}$) at a lag $l$ can be estimated as:
\begin{equation}
R(l) = \frac{K}{K-l} \frac{\sum_{i=1}^{K-l} (e(i)-\overline{e})(e(i+l)-\overline{e})}{\sum_{i=1}^{K} (e(i)-\overline{e})^2}
\label{autocorrfunc}
\end{equation} 
We calculate the autocorrelation series corresponding to $e(i)$ by applying ACF at various lags. Subsequently, the peaks (with height greater than a pre-defined value) in the autocorrelation series are found and the inter-peak gaps are calculated. If the variance of those gaps is lower than a threshold, the discrete sequence is detected as periodic. The complete algorithm proposed for detecting the presence of bot-CnC communication is shown in Algorithm \ref{A2}.
\begin{algorithm}[h]
	\centering	
	\caption{\scriptsize{Detect\_bot\_cnc\_comm (devind)}}
	\label{A2}
	\begin{algorithmic}[1]
		\State \textbf{INPUT}: devind (Index of IoT device under consideration)
		\State capfile $=$ \textsf{Capture\_traffic} (devind)
		\State capfile\_filter $=$ \textsf{Apply\_display\_filter} (capfile)
		\State cap\_seq $=$ \textsf{Encode\_pkt\_trace} (capfile\_filter)
		\State cap\_acf $=$ \textsf{Calculate\_ACF} (cap\_seq)
		\State peaks $=$ \textsf{Find\_peaks} (cap\_acf, height)
		\State peak\_diffs $=$ \textsf{Calculate\_peak\_gaps} (peaks)
		\If {\textsf{Variance} (peak\_diffs) $<$ \textit{thresh}}
			return (PERIOD\_DETECTED)
		\Else 
			\State return (PERIOD\_NOT\_DETECTED)
		\EndIf
	\end{algorithmic}
\end{algorithm}

\textbf{Handling False Positives/Negatives}: EDIMA is designed to handle false positives/negatives in the edge gateway’s aggregate traffic session classification. For example, let us assume that there is a false positive, i.e., a traffic session is classified as \textit{malicious} though it is \textit{benign}. The ML-based bot detector stage after ML classifier (Fig. \ref{EDIMA-bot-detect}) which detects the bot device(s) using ACF-based test, would not detect any bot and therefore EDIMA would know that there was a false positive in traffic session classification. However, assuming that there is a false negative, i.e., a traffic session is classified as \textit{benign} though it is \textit{malicious}, the ACF-based test stage would not be invoked. We propose to handle this by taking the average of classification results for a fixed number of consecutive traffic sessions subsequent to a session being classified as \textit{benign}. It is expected that by running over multiple sessions, the classification algorithm would be able to distinguish between true positives and false negative in more sessions than not and thus the averaging of classification results would lead to a \textit{true positive} classification.

\subsection{Policy Engine}
\label{policyengine}
Once a device or set of devices has been detected as bot(s), the policy engine decides the actions to be taken against those devices. We envision a set of default policies pre-configured by the network administrator along with the capability to make and update new policies as part of this module. For instance, some of the expressions for policy configuration can be:

\textit{policy-engine --create-policy <policy-name>}

\textit{policy-engine --add-action <policy-name> --dev <device-name> --action <action-name>}

\textit{policy-engine --delete-action <policy-name> --dev <device-name> --action <action-name>}

\textit{policy-engine --delete-policy <policy-name>}

We now mention some of the actions which can be taken by the policy engine. One such action where the bot traffic is blocked by the network administrator was mentioned in Section \ref{edimaarch}. Another alternative is to allow the bot to communicate with a few secure domains only for malware infection remediation \cite{ietfbot}. It is also possible to place the bot under continuous monitoring and deny all other communication except that required for the functioning of the underlying IoT device.

\section{Performance Evaluation}
\label{perfeval}
\subsection{Testbed Description}
\label{testbed}
To evaluate the performance of EDIMA on real devices, we built a testbed with commercial IoT and non-IoT devices which are listed in Table \ref{listofdevices}. 
Typical network applications running on the laptops/desktops were web browser (accessing web pages, video streaming sites, e.g., YouTube), email clients, etc.  Similarly, applications such as web browser, social media (Facebook/Twitter/LinkedIn), chat (WhatsApp), etc., were running on the smartphones along with a few other network applications in the background. The devices were used by 3 staff members in our lab over a period of 4 weeks, and thus the traffic data collected from those devices reflects real-world users' behaviour. 

\begin{table}[h]
	\centering
    \begin{tabular}{ | l | l | l |}
    \hline
    \textbf{Device} & \textbf{Brand} & \textbf{Type} \\ \hline
    Smart bulb & Philips Hue & IoT \\ \hline
    Smart bulb (LB100) & TP-Link & IoT \\ \hline
    IP camera (DCS-930L) & D-Link & IoT \\ \hline
    IP camera (DCS-5030L) & D-Link & IoT \\ \hline
    Smart Plug (HS110) & TP-Link & IoT \\ \hline
    Smart Switch (INSIGHT) & WeMo & IoT \\ \hline
    Smart Home Assistant & Google & IoT \\ \hline
    SmartThings Hub & Samsung & IoT \\ \hline
    Motion Sensor (DCH-S150) & D-Link & IoT \\ \hline
    Smart Plug (DSP-W215) & D-Link & IoT \\ \hline
    Laptop (Windows OS) & HP & Non-IoT \\ \hline
    Laptop (Windows OS) & Dell & Non-IoT \\ \hline
    Desktop (Ubuntu OS) & Dell & Non-IoT \\ \hline
    Smartphone (Android OS) & Samsung & Non-IoT \\ \hline
    Smartphone (Android OS) & One Plus & Non-IoT \\ \hline
    \end{tabular}
    \caption{List of IoT and Non-IoT Devices used in our Testbed}
    \label{listofdevices}
\end{table}

\begin{figure}[h]
\centering
	\begin{subfigure}[b]{0.4\textwidth}
		\centering	
		\includegraphics[width=\textwidth]{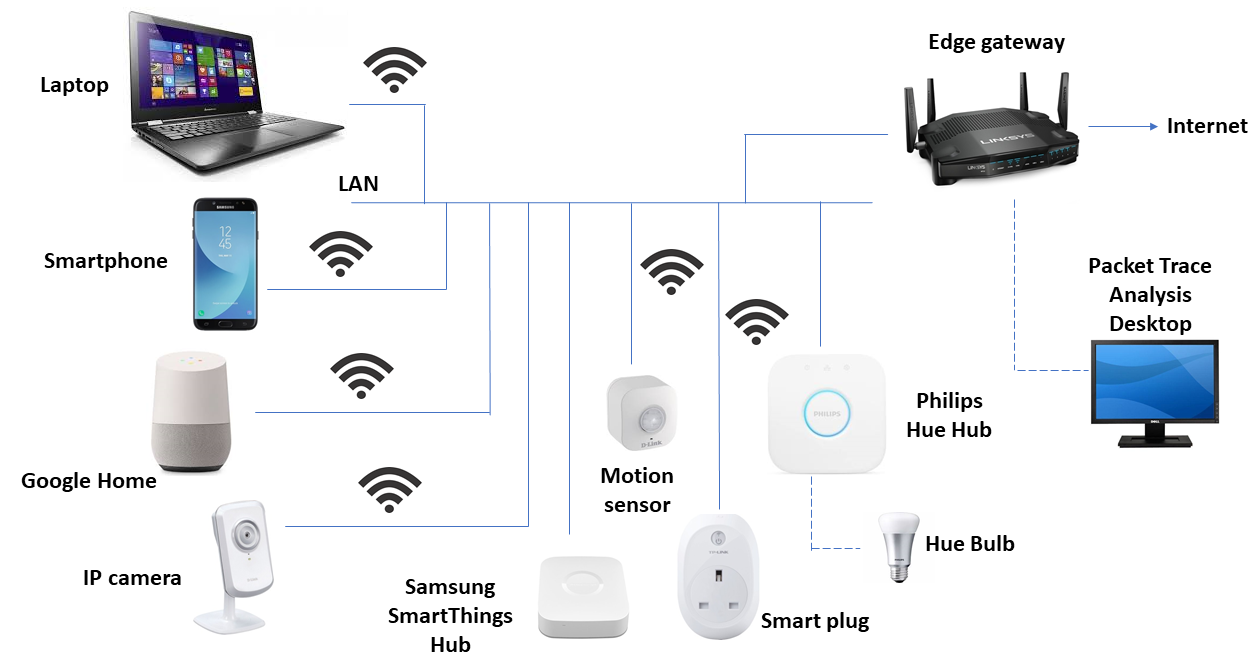}
		\caption{Schematic of the testbed}
		\label{EDIMA-testbed}
	\end{subfigure}
	\par\medskip
	\begin{subfigure}[b]{0.4\textwidth}
		\centering	
		\includegraphics[width=\textwidth]{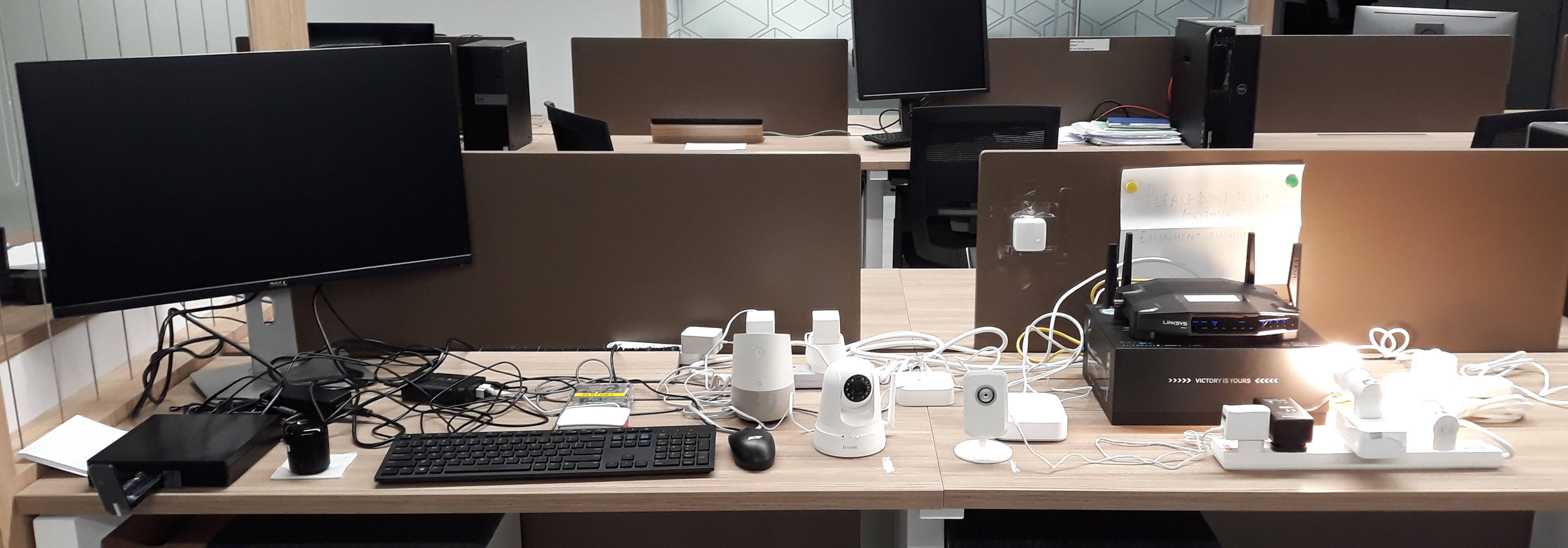}
		\caption{Picture of the actual testbed}
		\label{EDIMA-testbed-pic}
	\end{subfigure}
	\caption{Testbed used for performance evaluation of EDIMA}
\end{figure}

The edge gateway where the traffic from all the above devices was aggregated was a Linksys WRT32X router with a Marvell Armada 385 88F6820 1.8 GHz dual-core processor, 512MB RAM, 256MB NAND flash memory supporting IEEE 802.11a/b/g/n Wi-Fi standards. The Linksys WRT32X router comes with out-of-the-box OpenWRT \cite{openwrt} support (Version: Bleeding Edge). 
The packet traces collected from the router were analysed on a Dell OptiPlex 5060 desktop PC with an Intel Corei5-8500T 2.1GHz six-core processor, 8GB RAM running Ubuntu 18.04. 
The testbed schematic and picture are shown in Fig. \ref{EDIMA-testbed} and \ref{EDIMA-testbed-pic}.

\subsection{Data Collection}
\label{datacollect}
As mentioned in Section \ref{classifierdetails}, we classify aggregate traffic at the gateway into \textit{benign} or \textit{malicious}. Therefore, we need training data samples to represent both classes. Benign traffic can be generated through the normal operation of uninfected devices. However, malicious traffic consists of both benign traffic as well as malware scanning/infection packets. To generate malicious traffic, we have three options:
\begin{enumerate}
\item Run live malware on the device(s) in our testbed
\item Emulate the malware on our testbed
\item Replay live malware traffic on our testbed
\end{enumerate}

The first option, though capable of generating the most realistic malicious traffic, is ruled out as the live malware may infect other devices connected to the Internet and use them for attacks. Moreover, none of the IoT devices in our testbed had any known software vulnerabilities which could have been exploited by existing IoT malware. We used the second option in our earlier work \cite{ayush-wfiot}. However, emulating malware requires assumptions about their behaviour (such as scanning, communication with CnC server) which may not be fully representative of live malware. 

Therefore, in this work, we adopted the third option. We obtained 23 live IoT malware samples over a period of 3 months (May-July, 2019) from APIs provided by New Sky Security \cite{newskysecurity} and malware hosting server links posted by Bad Packets Report Twitter account \cite{badpackets}, henceforth referred to as the \textit{IoT-BPR-NSS} dataset. The FANTASM framework, provided by DeterLab \cite{deterlab} team for safe experimentation with live malware based on \cite{fantasm}, was used to run the malware samples and collect the packet traffic generated by them. 

Many of the malware samples were simple variants of each other as revealed by analysing their traffic using Wireshark. Those simple variants were discarded as they would not have added any value to our classification dataset. We ended up with two malware samples, called \textit{loligang} and \textit{echobot} by their authors, which belong to the TELNET and HTTP POST+GET categories (Appendix \ref{iot-malw-categ}) respectively. We could not obtain malware samples from other categories, as many of those malware were not active by the time we started collecting samples. We ran \textit{loligang} and \textit{echobot} binaries on the FANTASM testbed for 5 minutes each and captured the corresponding traffic \textit{pcap} files. Malicious traffic was then generated by replaying the malware traffic collected from FANTASM on the edge gateway using the \textit{tcpreplay} utility. We could not replay the malware traffic on commercial IoT devices in our testbed as they do not provide a shell to users. 

We used traffic session durations of 5, 10 and 15 minutes for this study.
1000 traffic sessions were captured for benign traffic and a further 1000 sessions for malicious traffic through our testbed. The malicious traffic sessions consisted of $400$ sessions corresponding to \textit{loligang}, another $400$ sessions corresponding to \textit{echobot} and the remaining $200$ sessions corresponding to both \textit{loligang} and \textit{echobot} traffic replayed at the OpenWRT router. The features mentioned in Section \ref{classifierdetails} were extracted from the captured sessions. Finally, appropriate class labels were assigned to the extracted feature vectors. 

Apart from the \textit{IoT-BPR-NSS} dataset, we also considered the \textit{IoT-23} dataset \cite{iot-23-dataset} which captured the traffic for 20 IoT malware samples from 2018 to 2019. However, roughly half the dataset consisted of Mirai or Mirai-like malware traffic, which we already had from the live malware samples we had obtained. The other half of the dataset consisted of more sophisticated centralized and P2P architecture botnet malware (e.g., Kenjiru, Okiru, Hakai, Hajime). The \textit{pcap} files corresponding to centralized architecture malware did not contain scanning packet traffic and therefore, we could not use their data in training EDIMA's ML algorithms for aggregate gateway traffic classification. The P2P botnet malware traffic files were not used as the focus of our work is on centralized botnet malware.

Both the \textit{IoT-23} and \textit{IoT-BPR-NSS} datasets consisted of bot-CnC server communication packets. We used them to verify our assumption that bot-CnC messaging is periodic as stated in Section \ref{botidentifier}. The malware \textit{pcap} files obtained from those datasets were analysed for bot-CnC messaging packets. This was done by filtering the the \textit{pcap} files for repetitive connections (including TCP ([PSH, ACK], [ACK]) or UDP messages) with a suspicious CnC server and checking the timings of those connections for periodicity.

 \begin{table}[h]
	\centering
    \begin{tabular}{ | p{1cm} | p{1cm} | p{3cm} | p{2cm} | }
    \hline
    \textbf{Dataset Name} & \textbf{Dataset Type} & \textbf{Purpose} & \textbf{Justification}\\ \hline
    \textit{IoT-NSS-BPR} & Malware samples & 1. Generate malicious aggregate traffic training data using FANTASM. 2. Verify assumption regarding periodicity of bot-CnC server messaging. & Consists of 23 live IoT malware samples. \\ \hline
    IoT-23 dataset & Malware traffic pcap files & Verify assumption regarding periodicity of bot-CnC server messaging only. & Consists of bot-CnC communication packet traces for IoT malware. \\ \hline
    UNSW IoT dataset & Aggregate IoT traffic pcap files & Test EDIMA’s robustness to the scaling of number of IoT devices connected to the edge gateway. & Provides \textit{pcap} files for aggregate traffic captured from 28 different uninfected IoT devices collected at a gateway. \\ \hline
    \end{tabular}
    \caption{Data sources used in our work}
    \label{listofdevices}
\end{table}   


\subsection{Data preparation}
The feature vectors obtained in the previous section for benign and malicious classes were checked for missing values and handled appropriately. Next, all the values in a feature vector were scaled using a MinMaxScaler \cite{scikitlearn} to lie within the range (0,1). This was done to avoid bias due to different scales of features. Further, the feature vectors were randomly permuted so that the sequence in which they were extracted does not affect the training of the classifier. The combined benign and malicious feature vectors were randomly divided into \textit{training} and \textit{test} datasets using an 80:20 split.

\subsection{Feature selection}
\label{feat-sel}
It is possible that some of the features extracted from training data are redundant or irrelevant with little to no contribution in improving the ML classifier's performance. Feature selection, or selecting a subset of relevant features, can help in simplifying the ML classifier model, shortening the time for training and reducing over-fitting. We used the $\chi^2$ statistical test to compute the $\chi^2$ test statistic for each feature from the sample data. The $\chi^2$ test values for features extracted from the 15 minutes traffic session dataset are shown in Table \ref{featscore1}.
Subsequently, we selected the best \textit{k}$=6$ features (having test statistic value more than zero) for training our ML classifiers. The selected features ranked by their $\chi^2$ test values are listed below:
\begin{enumerate}
\item Number of TCP half-open connections
\item Number of unique TCP SYN destination IP addresses
\item Minimum number of packets per unique destination IP address
\item Maximum number of packets per unique destination IP address
\item Mean TCP packet length
\item Mean number of packets per unique destination IP address
\end{enumerate}

\begin{table}
	\centering
    \begin{tabular}{ | l | l | }
    \hline
    \textbf{Feature} & \textbf{Score} \\ \hline
    Num\_half\_open\_conn & 165.130937 \\ \hline
    Num\_uniq\_Ipadd & 146.088020 \\ \hline
    Min\_pkt\_per\_Ipadd & 79.500000 \\ \hline
    Max\_pkt\_per\_Ipadd & 77.313488 \\ \hline
    Avg\_TCP\_pkt\_len & 13.700322 \\ \hline
    Avg\_pkt\_per\_Ipadd & 8.317776 \\ \hline
	Max\_TCP\_pkt\_len & 0 \\ \hline
	Min\_TCP\_pkt\_len & 0 \\ \hline
    \end{tabular}
    \caption{$\chi^2$ test scores for classifier features}
    \label{featscore1}
\end{table}


\subsection{PCA Analysis}
\begin{figure}[h]
	\centering
	\includegraphics[scale=0.2]{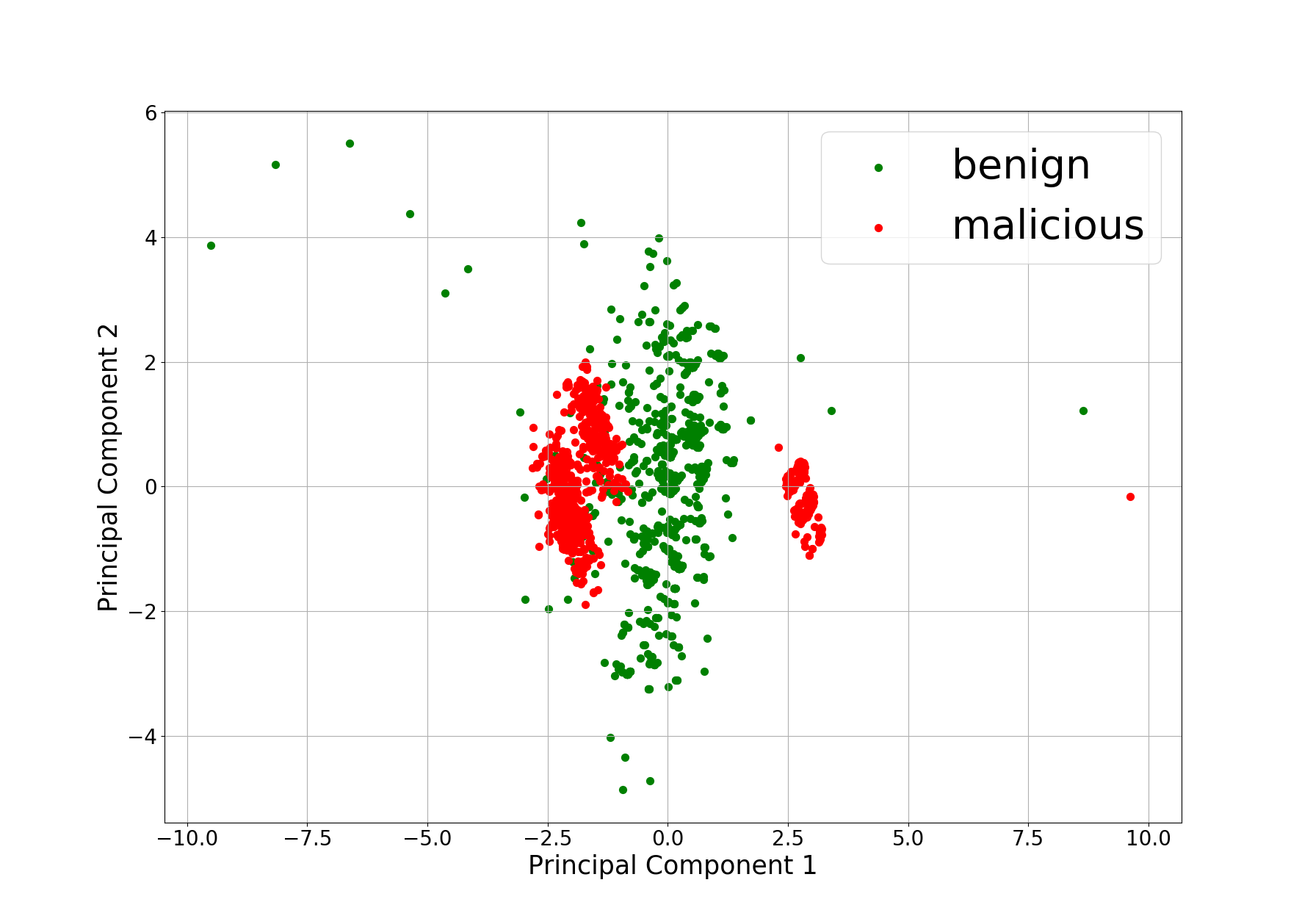}
	\caption{PCA analysis of the final prepared data with two components}
	\label{PCA-edima}
\end{figure}

In order to visualize the final data obtained in the previous sub-section and check if the benign and malicious data points are separable using the features proposed, we performed a Principle Component Analysis (PCA). PCA reduces the dimensionality and captures the big principal variance in a set of observations (feature vectors in our case) by generating new variables, called principle components (PCs). These PCs are orthogonal to each other and obtained as the eigenvectors of the largest eigenvalues of the covariance matrix of the original set of observations. A 2D scatter plot of two PCA components extracted from the final data for 15 minutes traffic session is shown in Fig. \ref{PCA-edima}. The benign and malicious PC points seem to be separable using a pair of decision boundaries on either side of the benign PC points cluster. There will still be a small number of outliers though, especially around the left decision boundary which does not provide a clear separation.

\subsection{Results}
\subsubsection{Scanning Activity Detection Performance}
\label{detect-perf}
We trained the following ML models using the final feature vectors obtained in Section \ref{feat-sel} after completing all the data processing steps: Gaussian Naive Bayes' (GNB), Support Vector Machine (SVM) and Random Forest (RForest). Subsequently, the trained ML models are used to predict the class labels of the test dataset and thereby, the detection performance of the models is evaluated and compared.
In this work, a 10-fold cross validation (CV) approach is used to tune the hyper-parameters of the ML classifiers for achieving the highest possible CV scores. The cross validation is based on training data only without using any information from the test dataset. The hyper-parameter values so obtained along with the associated CV scores for the 15 minutes traffic session data are displayed in Table \ref{hyperparams}.
%
%
%
\begin{table}[]
	\centering
  \begin{adjustbox}{width=\columnwidth}
    \begin{tabular}{ | l | l | p{.6\columnwidth} | }
    \hline
    \textbf{Classifier} & \textbf{CV Score} & \textbf{Hyper-parameter Values} \\ \hline
    GNB & 0.98 (+/- 0.02) & priors=None, var\_smoothing=1e-3 \\
 	SVM & 1.00 (+/- 0.01) & C=1.0, kernel='rbf', degree=3, gamma='scale', coef0=0.0, shrinking=True, probability=False, tol=0.001, cache\_size=200, class\_weight=None, verbose=False, max\_iter=-1, decision\_function\_shape='ovr', random\_state=None \\
 	RForest & 1.00 (+/- 0.00) & n\_estimators=10, criterion='gini', max\_depth=None, min\_samples\_split=2, min\_samples\_leaf=1, min\_weight\_fraction\_leaf=0.0, max\_features='auto', max\_leaf\_nodes=None, min\_impurity\_decrease=0.0, bootstrap=True, class\_weight=None \\ \hline 
    \end{tabular}
  \end{adjustbox}
    \caption{CV Scores and optimal hyper-parameter values for ML classifiers trained on testbed dataset (15 mins)}
    \label{hyperparams}
\end{table} 


Using the tuned hyper-parameters' values, the average classification accuracy (AC), precision (PR), recall (RC) and F1 scores obtained for the final classifiers over 50 runs are shown in Table \ref{class-scores}. It can be observed that the Random Forest classifier performs the best in terms of classification accuracy followed by SVM classifier and Gaussian Naive Bayes' classifier. All the classifiers report a \textit{precision} of close to $1.0$, i.e., an extremely low false positive rate, which is a much desired property for bot detection. This is because we do not want benign IoT devices to be classified as bots which could lead to further administrative actions on those devices. Moreover, all the classifiers report a \textit{recall} of close to $1.0$ as well, i.e., an extremely low false negative rate, which is also a much desired property for bot detection. False negatives lead to bots getting classified as benign IoT devices and as a result, the network administrator could overlook them while the bots continue to pose a security threat without being detected. Based on the classification performance obtained above, we use the Random Forest model for aggregate traffic classification while implementing EDIMA in code.

\textit{Performance Comparison With Previous Works}: The performance of the proposed aggregate traffic classifier for scanning activity detection is compared with the botnet scanning traffic detector introduced in \cite{botscantraffic1} here. We desist from comparing with \textit{Kitsune}\cite{kitsune} as it is an anomaly detection system whereas EDIMA is based on a misuse-detection approach. The two approaches are quite different from each other \cite{intrusion-det-survey} with anomaly detection systems typically performing worse in terms of false positive rates. Further, \textit{Kitsune} uses per-packet classification whereas EDIMA performs per-session classification. If we wish to use per-session classification with \textit{Kitsune}, we would have to make significant changes in its source code. Also, it is not possible to use per-packet classification with EDIMA as it is designed to use session-based features due to the reasons explained in Section \ref{classifierdetails}. Other anomaly detection-based systems for IoT \cite{mud-SDN-ID, nbaiot, diot} are also excluded from performance comparison due to the reason explained above as well as lack of publicly available source code.

%
We split our benign and malicious traffic datasets into training and test datasets using a 80:20 ratio, which is the split used during the evaluation of our ML-based aggregate traffic classification algorithm. We construct $H_m$, the first conditional entropy vector using samples from \textit{malicious} training data and $H_t$, the second conditional entropy vector using a test sample and samples from \textit{malicious} training data. The conditional entropies are calculated using NPEET Python package \cite{npeet}. Using the KS test, we attempt to decide whether the test sample (or traffic session) is similar to malicious traffic in terms of conditional entropy distribution to classify the test sample as malicious. A low false positive probability, $\gamma = 0.05$ is selected, which gives the test threshold $K_\gamma$ as $0.51961$.

In applying KS test for classification of test samples, we define \textit{true positive} (TP) as the event in which a malicious test sample is found similar to malicious training dataset and \textit{false positive} (FP) when a benign test sample is found similar to malicious training dataset. When the same malicious test dataset (consisting of 200 samples) was tested for similarity with malicious and benign training datasets, the number of TPs and FPs were calculated as $186$ and $168$ respectively. It can be inferred that few test samples were found similar to both malicious and benign traffic which means that conditional entropy based KS test can not be used for reliable classification. Further, the malicious traffic detection accuracy (or \textit{precision} in ML terminology) can be calculated as $186/200 = 93\%$, which is lower than the precision of all the ML classifiers as shown in Table \ref{class-scores}. Thus, based on accuracy and reliability of classification, we can conclude that our ML-based aggregate traffic classification method performs much better than the conditional entropy based KS test in \cite{botscantraffic1}.

\begin{table}[]
	\centering
	\begin{tabular}{|l|c|l|l|l|l|l|}
	\hline
	\textbf{DATASET} & \multicolumn{1}{l|}{\thead{\textbf{SESSION}\\ \textbf{DURATION}}} & \textbf{MODEL} & \textbf{AC} & \textbf{PR} & \textbf{RC} & \textbf{F1} \\ \hline
	\multirow{9}{*}{Testbed} & \multirow{3}{*}{5 mins} & Rforest & 1.0 & 1.0 & 1.0 & 1.0 \\
	 &  & SVM & 0.99 & 0.99 & 1.0 & 0.99 \\
	 &  & GNB & 0.95 & 0.95 & 0.97 & 0.96 \\ \cline{2-7}  
	 & \multirow{3}{*}{10 mins} & Rforest & 1.0 & 1.0 & 1.0 & 1.0 \\ 
	 & 	& SVM & 0.99 & 0.99 & 1.0 & 0.99 \\
	 &  & GNB & 0.96 & 0.96 & 0.98 & 0.97 \\ \cline{2-7} 
	 & \multirow{3}{*}{15 mins} & Rforest & 1.0 & 1.0 & 1.0 & 1.0 \\
	 &  & SVM & 0.99 & 0.99 & 1.0 & 0.99 \\
	 &  & GNB & 0.97 & 0.97 & 1.0 & 0.97 \\
	 &  & KS Test & - & 0.93 & - & - \\ \hline
	\multirow{3}{*}{UNSW} & \multirow{3}{*}{15 mins} & Rforest & 1.0 & 1.0 & 1.0 & 1.0 \\
	 &  & SVM & 0.91 & 0.91 & 0.79 & 0.88 \\
	 &  & GNB & 0.92 & 0.92 & 0.83 & 0.91 \\ \hline
	\end{tabular}
	\caption{Performance of ML Classifiers for Scanning Activity Detection}
	\label{class-scores}
\end{table}


\subsubsection{Bot-CnC Communication Detection Performance}
To evaluate the performance of our proposed $Detect\_bot\_CnC\_comm$ algorithm, we first consider a scenario where we we use the malicious traffic dataset (with a 15 minutes session duration) generated in Section \ref{datacollect} by replaying the \textit{loligang} pcap file. We know from the ground truth that the \textit{loligang} pcap file contains bot-CnC messaging packets with an approximate periodicity of $60$ seconds. Each \textit{pcap} file in the malicious traffic dataset is processed as follows:
\begin{enumerate}
	\item The packet traffic associated with each connected IP address (corresponding to a device) detected by the gateway was separated from the aggregate traffic. 
	\item The devices' packet streams were filtered, sampled, encoded and mean subtracted as described in Section \ref{botidentifier} to produce discrete-time sequences. 
\end{enumerate}
We evaluate the detection performance of Algorithm \ref{A2} on the discrete-time sequences so obtained in terms of detection rate (DR) and missed-detection rate (MDR). \textit{Detection Rate} is the fraction of the total number of malicious traffic \textit{pcap} files which have been correctly detected as containing bot-CnC traffic, and \textit{Missed-detection Rate} is the fraction of the total number of malicious traffic \textit{pcap} files which have been incorrectly detected as not containing bot-CnC traffic.  

In the second scenario, we use the malicious traffic dataset (with a $15$ minutes session duration) generated in Section \ref{datacollect} by replaying the \textit{echobot} pcap file. We know from the ground truth that the \textit{echobot} pcap file contains bot-CnC messaging packets with an approximate periodicity of $210$ seconds. 
Using the parameter values (given in Table \ref{algo2params}), the detection performance of Algorithm \ref{A2} for both the scenarios specified above is shown in Table \ref{botcncdet-scores}. It can be seen that our algorithm gives a DR of $1.0$ and a MDR of $0.0$ for both the scenarios outlined above. 

The sub-sampling frequency of $0.1$ was selected to keep the number of samples to be processed by our algorithm and hence the processing time within a reasonable limit. The minimum autocorrelation peak height was kept at $0.7$ times the maximum autocorrelation peak height (excluding the peak at lag $0$) because any value above that corresponds to a significant degree of autocorrelation. The inter-peak gap variance threshold was kept at a low value of $0.01$ so that only when the autocorrelation function's peaks are almost equally separated is the underlying discrete-time sequence detected as periodic.

\begin{table}[h]
	\centering
    \begin{tabular}{ | l | l |}
    \hline
    Traffic sampling frequency & 0.1 \\ \hline
    Min. autocorrelation peak height & 0.7$\times$(Max. peak height) \\ \hline
    Inter-peak gap variance threshold & 0.01 \\ \hline
    \end{tabular}
    \caption{Parameter Values for Evaluation of Bot-CnC Communication Detection Performance}
    \label{algo2params}
\end{table}

\textit{Performance Comparison With Previous Works}: We evaluate the performance of the proposed $Detect\_bot\_CnC\_comm$ algorithm against the botnet detection test developed in \cite{botnetdetwalker} here. A performance comparison with BotHunter/BotSniffer/BotMiner \cite{bothunter, botsniffer, botminer} is not provided as those works detect CnC communication at the network level using correlation-based techniques whereas EDIMA's bot-CnC communication detection algorithm runs on the edge gateway directly connected to IoT devices where the above correlation techniques cannot be applied. Further, the source code for those works has not been released, making it difficult for us to replicate their proposed botnet detection systems.

We use the discrete-time sequences produced during the performance evaluation of our proposed $Detect\_bot\_CnC\_comm$ algorithm. In order to confirm that those sequences correspond to the bot-CnC communication traffic as per \cite{botnetdetwalker}, we apply the Walker's largest sample test (WLST) to the sequences. Keeping the false positive probability, $\gamma$ at a low value of $0.1$, the test gives a DR of $0.0$ and a MDR of $1.0$ for a session duration of 15 minutes, which is extremely poor compared to a DR of $1.0$ and MDR of $0.0$ given by our proposed $Detect\_bot\_CnC\_comm$ algorithm. This result was somewhat expected as the Walker's largest sample test requires a large number of time samples to be effective and we had only a limited number of time samples collected in the capture duration. As mentioned earlier in Section \ref{botidentifier}, in a real-world implementation, it is not recommended to have long capture durations as that would lead to a large delay in bot detection.

\begin{table}[]
	\centering
	\begin{tabular}{|l|c|l|l|l|}
	\hline
	\textbf{DATASET} & \multicolumn{1}{l|}{\thead{\textbf{SESSION}\\ \textbf{DURATION}}} & \textbf{METHOD} & \textbf{DR} & \textbf{MDR} \\ \hline
	\multirow{3}{*}{Testbed (loligang)} & \multirow{3}{*}{15 mins} & Algo 2 & 1.0 & 0.0 \\
	 &  & WLST & 0.0 & 1.0 \\  \hline
	 \multirow{3}{*}{Testbed (echobot)} & \multirow{3}{*}{15 mins} & Algo 2 & 1.0 & 0.0 \\
	 &  & WLST & 0.0 & 1.0 \\  \hline
	\end{tabular}
	\caption{Bot-CnC Communication Detection performance of Algorithm \ref{A2}}
	\label{botcncdet-scores}
\end{table}


\subsubsection{Robustness Test}
The number of IoT devices connected to a gateway needs to be scaled up to test EDIMA's robustness and study the effect on its performance. Towards this, we used the UNSW IoT dataset in \cite{unswdataset} which provides \textit{pcap} files for aggregate traffic captured from 28 different uninfected IoT devices collected at a gateway over 24 hours for 20 days. Half of the original \textit{pcap} files were used in generating benign traffic. To generate malicious traffic, we replayed the other half of \textit{pcap} files at the OpenWRT router in our testbed (with all IoT/non-IoT devices disconnected) in addition to a single \textit{loligang} \textit{pcap} file using the malware traffic replay technique mentioned in Section \ref{datacollect}. We captured the packet traffic at the router and followed the same procedure as used for our testbed dataset to train and test ML classifiers. 

The aggregate traffic classification performance of EDIMA's ML-based bot detector using the UNSW IoT dataset for 15 minutes session duration is presented in Table \ref{class-scores}. Again, the Random Forest classifier performs the best with accuracy and precision scores the same as that for our testbed dataset. However, SVM and Gaussian Naive Bayes' classifiers register a significant drop in accuracy and precision scores compared to their performance on our testbed dataset. Thus, after more than doubling the number of IoT devices from $11$ to $29$ with just one bot present among the IoT devices, ML classifiers such as Random Forest are still able to accurately detect malicious aggregate traffic with almost zero false positives.

Theoretically, up to 255 devices can be connected to a consumer edge gateway (e.g., Wi-Fi router or access point). The actual number of devices connected though is usually a lot lower and depends on the type of network in which the gateway is installed, e.g., home, campus, enterprise, with home networks (focus of our work) typically consisting of the lowest number of connected devices out of the three network types. As long as we are targeting current IoT malware whose scanning behaviour is quite similar to the malware tested, it is fair to claim that EDIMA will perform well even when the number of devices connected to a gateway goes up.

\subsubsection{Runtime Performance of EDIMA}
We have implemented EDIMA in Python\footnote{The source code has not been publicly released due to the terms of our project funding.}. Most of the commercially available consumer gateways/routers run proprietary firmware on top of Linux with basic utilities (e.g., Busybox) and a simple web interface. They do not support Python environment, DB services or Linux/Python package management which are required for running EDIMA code. Therefore, it is not possible to run the code directly on all consumer routers as is. Instead, we tested the code on a Raspberry Pi 3B+ (RPi) device with a Broadcom BCM2837B0 1.4GHz quad-core processor, 1GB SDRAM supporting IEEE 802.11.b/g/n/ac wireless LAN which is similar in configuration to a consumer Wi-Fi router. The test setup remains similar to our testbed described in Section \ref{testbed}, only with the OpenWRT router replaced by RPi and the exclusion of few devices which require wired Internet connectivity (e.g., Samsung SmartThings, Philips Hue bridge). The session duration used for traffic capture is $5$ minutes. We considered three performance quantifiers: \textit{Throughput}, \textit{Delay} and \textit{Memory statistics}. 

\textit{Throughput} refers to the average number of packets per second that are successfully processed. Now, EDIMA is an offline detector, so the number of packets processed per second by it are not as important as in the case of an online detector. Further, the percentage of packets dropped (which get excluded from the bot detection process) is completely determined by the packet capture function used in EDIMA's implementation which depends on the capture speed of the network card installed on the gateway where EDIMA is deployed. Therefore, we do not include throughput results in the final runtime performance evaluation.

Two types of \textit{Delay} can be calculated in the context of EDIMA: 
\begin{enumerate}
\item average time interval (in \textit{seconds}) between the arrival of the first malicious packet at the router and the detection of malicious traffic by EDIMA code ($T_{AD}$), and
\item average time interval (in \textit{seconds}) between the detection of malicious traffic by EDIMA code and the detection of the first bot responsible for malicious traffic ($T_{BD}$). 
\end{enumerate}
\textit{Memory statistics} refers to the amount of local router (flash) storage required for storing EDIMA code and the average RAM consumed when the code is running. The averaging in all the relevant performance metrics is done over 50 code runs each. The evaluation results are shown in Table \ref{edima-runtime}. The results point towards a system which is quite reasonable in terms of detection delay (considering that EDIMA is designed to be an offline detector), requires minimal storage space and consumes only a small fraction of the total RAM on an average (9.62\%).

\begin{table}[h]
	\centering
    \begin{tabular}{ | l | l |}
    \hline
    $T_{AD}$ & 415.6 secs \\ \hline
    $T_{BD}$ & 342.4 secs \\ \hline
    Local Storage & 12kB \\ \hline
    Avg. RAM & 96.2MB \\ \hline
    \end{tabular}
    \caption{Runtime performance of EDIMA on a Raspberry Pi 3B+}
    \label{edima-runtime}
\end{table}


\subsection{Limitations and Future Work}
\label{limitations}
As the Malware PCAP Database consists of \textit{pcap} samples from only those IoT malware which have been observed infecting honeypots, traffic generated by zero-day malware is not a part of the database and therefore, EDIMA may not be able to detect bots infected by those malware. Also, EDIMA uses supervised ML algorithms which means that it can detect only known malicious traffic patterns. However, supervised ML algorithms are used much more commonly compared to unsupervised algorithms in real-world anti-virus and anti-malware products to detect virus/malware behaviour as well as in actual deployments of Network Intrusion Detection Systems \cite{MLIDissues}.

Advanced malware which attempt to evade detection by EDIMA may slow down the bot scanning activity to confuse the trained ML algorithms classifying aggregate gateway traffic and lower its detection accuracy and precision. However, as mentioned earlier in this paper, this approach decreases the scanning efficiency of the malware resulting in the slowing down of malware's propagation through a network and may not be desirable for the malware author. Further, this evasion technique can be countered by increasing the traffic session duration to capture enough scanning packets though it may lead to longer classification delays. 

Evasive malware may also use some other bot-CnC server messaging mechanism than the ([PSH,ACK], [ACK]) one to escape filtering and/or force the bot-CnC communication to be non-periodic (by adding noise traffic for example). If the attacker does change the bot-CnC messaging protocol, the detection method can be changed accordingly. Moreover, our bot-CnC server communication detection algorithm uses an ACF-based approach which can detect periodicity in the presence of noise as well if the noise is uncorrelated with the desired signal (discrete-time sequence extracted from bot-CnC server traffic). Forcing the bot-CnC communication to be non-periodic may result in the CnC server losing control of the bot in between the keep-alive messages and cause the CnC server to have an unreliable estimate of the number of active, connected bots. This may lead to a decrease in the impact of any future attacks launched by the CnC server using its bots.

Adversarial machine learning may also be deployed by malware authors by changing the bot traffic in subtle ways so as to avoid detection by EDIMA's trained ML algorithms. Additionally, it is possible that some IoT devices connected to a gateway may already be infected before EDIMA is deployed and starts running on the gateway. As part of our future work, it may be interesting to explore ways of detecting and filtering traffic samples containing traffic from existing bots before fully deploying EDIMA. One possible approach is to use a reliable statistical test for checking if the captured traffic samples are similar to known malicious traffic samples in some probability measure. Finally, the aggregate traffic classifier trained by ML Model Constructor may have to be re-trained due to a number of reasons (mentioned in Section \ref{edimaarch}) and this may cause a delay in bot detection.


\section{Conclusion} 
We have proposed EDIMA, a lightweight solution for early detection of IoT botnets in home networks. It detects bots connected to an edge gateway in two stages- first by looking for scanning and subsequently bot-CnC server communication traffic patterns. EDIMA consists of a traffic parser, feature extractor, ML-based bot detector, policy engine, ML model constructor and a malware PCAP database. The operation of all the above components was discussed in detail. A comprehensive performance evaluation of EDIMA using our testbed setup revealed that it has a close to $100\%$ accuracy and a very low false positive rate in detecting malicious aggregate gateway traffic with ML algorithms such as the Random Forest. The detection performance holds even when the number of devices is more than doubled, suggesting that EDIMA's performance is robust to an increase in the number of IoT devices connected to the gateway. We also compared EDIMA's performance with existing techniques for bot scanning traffic and bot-CnC communication detection and found that EDIMA performs much better and more reliably. Finally, we analysed the runtime performance of a Python implementation of EDIMA on a Raspberry Pi which showed low bot detection delays and low RAM consumption.

\section*{Acknowledgment}
The authors would like to appreciate the National Cybersecurity R\&D Lab, Singapore for allowing us to use their testbed to collect important data which has been used in our work. This research is supported by the National Research Foundation, Prime Minister’s Office, Singapore under its Corporate Laboratory@University Scheme, National University of Singapore, and Singapore Telecommunications Ltd.

\bibliographystyle{ieeetran}
\begingroup
\raggedright
\bibliography{oqeprop2}
\endgroup

\appendices

\section{Background On IoT Botnets}
\label{background}
\subsection{IoT Malware Categorization}
\label{iot-malw-categ}
We have categorized known IoT malware based on type of protocol vulnerability that they target: TELNET, HTTP POST and HTTP GET. TELNET is an application-layer protocol used for bidirectional byte-oriented communication. Typically, a user with a terminal and running a TELNET client program, accesses a remote host running a TELNET server by requesting a connection to the remote host and logging in by providing its credentials. HTTP GET and POST are methods based on HTTP (HyperText Transfer Protocol) application-layer protocol which are used to request data from and send data to servers, respectively. For example, HTTP GET is commonly used for requesting web pages from remote web servers through a browser. 
The proposed malware categories, various malware belonging to those categories and brief descriptions of their operation are presented in Table \ref{table-malware}.
\begin{table}[]
	\centering
  \begin{adjustbox}{width=\columnwidth}
    \begin{tabular}{ | l | l | p{.6\columnwidth} | }
    \hline
    \textbf{Category} & \textbf{Malware} & \textbf{Description} \\ \hline
    \multirow{6}{*}{TELNET} & Mirai & Sends SYN packets to probe open TELNET ports at random IP addresses. If successful, it tries to login using list of default credentials. \\
 	& Hajime & Same propagation mechanism as Mirai, but no CnC server. Instead, it is built on a P2P network. Purpose seems to be to improve security of IoT devices.  \\
 	& Remaiten & Same propagation mechanism as Mirai. Downloads binary specific to targeted platform. Uses IRC protocol for CnC server communication.\\
 	& Linux.Wifatch & Same propagation mechanism as Mirai. Apparently, it tries to secure IoT devices from other malware.  \\
 	& Brickerbot & Rewrites the device firmware, rendering the device permanently inoperable. \\
 	& Torii & Sophisticated unlike Mirai-variants and uses a number of advanced techniques such as rich set of features to extract sensitive information, modular architecture and encrypted communications. \\ \hline 		\multirow{4}{*}{HTTP POST} & Satori & Sends NewInternalClient request through miniigd SOAP service (REALTEK SDK) or sends malicious packets to port 37215 (Huwaei home gateway). \\
 	& Masuta & Forms SOAP request which bypasses authentication and causes arbitrary code execution.  \\
 	& Linux.Darlloz & Sends HTTP POST requests by using  PHP 'php-cgi' Information Disclosure Vulnerability to download the worm from a malicious server on an unpatched device.  \\
 	& JenX & Exploits SOAP based vulnerabilities in Huawei HG532 routers, D-Link devices (running HNAP protocol) and Realtek SDK miniigd service; contains option for DDoS attack. \\ \hline
 	\multirow{1}{*}{HTTP GET} & Amnesia & Makes simple HTTP requests, searches for a special string “Cross Web Server” in the HTTP response from target. If successful, sends four more HTTP requests which contain exploit payloads of four different shell commands.  \\ \hline 	
 	\multirow{1}{*}{TELNET + HTTP GET} & Hide-n-Seek & Sends HTTP request embedded with malicious command to vulnerable devices (AVTECH IP cameras, NVR, DVR, Wansview IP camera), first IoT malware to persist across device reboots. \\ \hline 
 	\multirow{2}{*}{HTTP POST + GET} & Reaper & Scans first on a list of TCP Ports to fingerprint devices, then second wave of scans on TCP ports running web services such as 80, 8080\dots, sends HTTP POST or GET request for command injection, usually through CGI or PHP.\\
 	& Echobot & Based on Mirai source code and targets 26 different RCE exploits affecting NAS, routers, NVRs, IP cameras etc. \\ \hline 
 	\multirow{1}{*}{TELNET + HTTP POST} & Hakai & Based on Qbot (an old IoT malware strain), targets similar exploits as JenX in addition to employing a TELNET scanner.  \\ \hline
 	\multirow{1}{*}{UPnP} & BCMUPnP\_Hunter & Exploits an old vulnerability in Broadcom's UPnP (Universal Plug and Play) protocol which is used in millions of routers including D-Link, TP-Link, ZTE etc; uses a self-built proxy network which communicates with email providers, most probably for spamming; employs a complex multi-step infection process.  \\ \hline
    \end{tabular}
  \end{adjustbox}
    \caption{IoT Malware Categories}
    \label{table-malware}
\end{table}

\section{Bot Detection Confidence Score} 
\label{bdcs}

Let $P(Y_{PD}^i | X_{MC})$ be the conditional probability of detecting periodicity due to bot-CnC communication in the $i^{th}$ device's traffic (using the ACF-based algorithm \ref{A1}) once the aggregate traffic has been classified as malicious. It is defined as follows:
\begin{equation}
\label{period-det-prob}
P(Y_{PD}^i | X_{MC}) \approx
	\begin{cases}
 		1, & \text{if p-value at largest lag}\ \ll \alpha \\
		Pr (Q > \chi^2_{1-\alpha, h}), & \text{otherwise}
	\end{cases}
\end{equation} 
where $Q$ is the Ljung–Box test statistic \cite{ljung-box}, $\alpha$ is the significance level, \textit{p-value} is the probability that the test statistic assumes a value equal to or greater than the observed value, and $\chi^2_h$ is the Chi-square distribution with $h$ degrees of freedom. The Ljung–Box test is a binary hypothesis test \cite{stevenkaydet} which decides if a time series exhibits a significant amount of combined autocorrelation at specified number of lags ($h$ in this case) or if there is zero autocorrelation at those lags. As $Q$ approaches a $\chi^2_h$ distribution asymptotically, $P(Y_{PD}^i | X_{MC})$ is only bounded by $Pr (Q > \chi^2_{1-\alpha, h})$ and therefore, the approximation in Equation \ref{period-det-prob} is not perfect. However, it serves the purpose as otherwise, we have no way of approximating the detection performance of the ACF-based algorithm \ref{A1}.

Hence, the BDCS (Bot Detection Confidence Score) for EDIMA, which represents the likelihood/confidence level of the accurate detection of bots among the IoT devices under test, is hereby defined as:
\begin{equation}
\begin{split}
BDCS &= P (Y_{PD} = 'bot' | X_{MC} = 'malicious') \\
&= \prod_{i=1}^N P (Y_{PD}^i = 'bot' | X_{MC} = 'malicious')
\end{split}
\end{equation}
Assuming total $N$ number of IoT devices, ($P (Y_{PD} = 'bot' | X_{MC} = 'malicious')$) is written as a product of probabilities ($P (Y_{PD}^i = 'bot' | X_{MC} = 'malicious')$), because we assume that all the devices are likely to be infected independent of each other (which is true in case of most IoT malware as they scan randomly generated IP addresses).

\end{document}